\documentclass[journal]{IEEEtran}

\usepackage{amsmath, amsfonts, amssymb, cite, epsfig}
\usepackage{url}
\usepackage[breaklinks]{hyperref}
\usepackage{breakurl}

\def \1{\mathbf{\,1}}

\graphicspath{{figures/}}

\def \FigWidthSmallEr{0.35\textwidth}
\def \FigWidthSmall{0.43\textwidth}
\def \FigWidth{0.48\textwidth}

\begin{document}

\title{Cyber Deception for Computer and Network Security: Survey and Challenges}

\author{Zhuo Lu,~\IEEEmembership{Member,~IEEE,}
        Cliff Wang,~\IEEEmembership{Fellow,~IEEE,}
        and Shangqing Zhao~\IEEEmembership{Student Member,~IEEE,}
\thanks{Zhuo Lu and Shangqing Zhao are with the Department of Electrical Engineering, University of South Florida, Tampa FL 33620, USA. E-mails: \{zhuolu@, shangqing@\}usf.edu.}
\thanks{Cliff Wang is with the Department of Electrical and Computer Engineering, North Carolina State University, Raleigh NC 27695, USA. Email: cliffwang@ncsu.edu.}
}

\maketitle

\begin{abstract}

Cyber deception has recently received increasing attentions as a promising mechanism for proactive cyber defense. Cyber deception strategies aim at injecting intentionally falsified information to sabotage the early stage of attack reconnaissance and planning in order to render the final attack action harmless or ineffective.

Motivated by recent advances in cyber deception research, we in this paper provide a formal view of cyber deception, and review high-level deception schemes and actions. We also summarize and classify recent research results of cyber defense techniques built upon the concept of cyber deception, including game-theoretic modeling at the strategic level, network-level deception, in-host-system deception and cryptography based deception. Finally, we lay out and discuss in detail the research challenges towards developing full-fledged cyber deception frameworks and mechanisms.
\end{abstract}

\section{Introduction}
As the convergence of our physical and digital worlds grows quickly, more and more information becomes available and it is a critical task to protect today's information technology (IT) systems and the information they carry. Recent high-profile hacking events (e.g., 2014 Sony Pictures hack \cite{sony-hack} and 2016 Democratic National Committee email hack \cite{DNC-hack}) and steadily increasing statistics of cyber attackers \cite{TP-toolbox-web} have shown that our current cyber defense is inadequate to combat more prevalent and sophisticated cyber attacks.

From an attacker's perspective, a cyber kill chain \cite{croom2010cyber, zaffarano2015quantitative} consists of a number of attack stages, from the reconnaissance stage to understand a computer network system towards the final stage to launch effective attacks. Traditional cyber defense mechanisms for computer and network systems have been largely reactive in nature, and have extensively addressed attack detection and mitigation after the attack action, but offers ineffective countermeasures at early stages in the cyber kill chain.

Recently, proactive strategies have been proposed to counter cyber attacks in their early stages. Such strategies can be mainly categorized into two types: 1) moving target defense (MTD, e.g., \cite{jafarian2012openflow, 15cpf-mtd, zaffarano2015quantitative, carroll2014analysis, jajodia2011moving} and 2) cyber deception (e.g., \cite{stoll1989cuckoo, 7070445, spitzner2003honeypots, juels2013honeywords}). With the same objective to defeat attacks, MTD and cyber deception entail different defense procedures:MTD focuses on dynamically changing the attack surface (e.g., system setups or configurations) such that an attacker cannot observe and identify accurate information during the attack reconnaissance stage, which can make any attack action hardly effective. Cyber deception aims at injecting intentionally falsified information to mislead the attackers during their attack planning stage. For example, the early honeypot systems \cite{spitzner2003honeypots, 7070445} intend to attract potential attackers to waste their time while in the meantime trying to learn attackers' strategies.

Compared to MTD, cyber deception has several advantages and can be more effective when used appropriately \cite{17wl-sp}. There are a number of cyber deception based strategies created in the literature to protect computing or network systems, showing the promising potential of cyber deception to be adopted as a mainstream proactive cyber defense technique. Motivated by the recent research progress, we aim to take a formal look at cyber deception, and review common deception schemes and actions. In particular, the following issues are discussed in this paper.

\begin{itemize}
\item Top-level model of cyber deception: we first describe the stages of one-round deception model, and then review the common deception schemes and actions.
\item State-of-the-art: we summarize recent research results of cyber defense techniques built upon the concept of cyber deception in the literature, including game-theoretic modeling at the strategic level, network deception approaches, in-host-system deception schemes and finally cryptography based deception methods.
\item Major research issues and challenges: we identify major research issues associated with cyber deception, and discuss each issue in detail.
\end{itemize}

The rest of the paper is as follows. In Section~\ref{Sec:Models}, we review the top-level models of cyber deception. In Section~\ref{Sec:CurResearch}, we categorize recent research results leveraging cyber deception in the literature. In Section~\ref{Sec:Directions}, we outline and discuss the major research issues going forward. Finally, we conclude this paper in Section~\ref{Sec:Con}.

\section{High-level Models of Cyber Deception}\label{Sec:Models}
In this section, we review the high-level models of cyber deception. We first discuss the three phases of one round of deception, and summarize common deception schemes and actions.

\subsection{Model of One-Round Deception}
Cyber deception strategies are ``planned actions taken to mislead and/or confuse attackers and to thereby cause them to take (or not take) specific actions that aid computer security defenses.'' \cite{yuill2006defensive}. In practice, an attacker may keep probing to a target system to find potential vulnerability to exploit, while a defender may choose to update its system regularly to defeat reconnaissance and potential exploits.  The defense framework based on cyber deception can be modeled as a two-party interactive process. Attacker and defender engagement consistently evolves over time. Defense using cyber deception may never be a single action but quite often involve multiple rounds of engagement to be effective.

\begin{figure}[h]
    \centering
    \includegraphics[width=\FigWidthSmall]{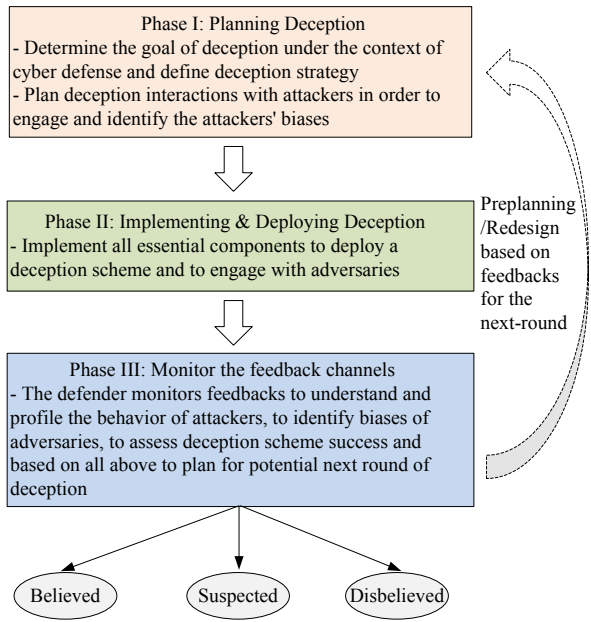}
    \caption{One-round model of deception defense.}
    \label{Fig:OneRoundModel}
\end{figure}

For each round of cyber deception, it may include three-phase actions \cite{14as-nspw, heckman2015denial}, as shown in Figure~\ref{Fig:OneRoundModel}.

\begin{itemize}

\item Phase I (planning deception). During the planning phase of deception, based on the initial knowledge of adversaries such as their intents, interests and capabilities, defenders would first specify the goal of the deception that could reasonably be expected as achievable. A deception plan would include the design on how to engage with the attackers and the deception information that could be used to create cognitive biases \cite{17wl-sp}, which is the key to the success of the overall deception-based defense.

The defender would carefully balance the details of the deception plan (e.g., the amount and type of truth disclosure, combined with biased information in order to deceive, potential risks to confuse regular users, and the overhead involved) to maximize the success rate of the deception strategy while trying to minimize the impact to normal operations.

\item Phase II (implementing and deploying deception). In this phase, the deception scheme is implemented. Depending on the deception plan, deception components could contain both devices (hosts or servers), information on devices, and communications among devices. For more advanced adversaries, defenders may also need to pay attention to side channels so that the deployed deception scheme will be as foolproof as possible.

\item Phase III (monitoring and evaluating deception outcome) is the final phase of one round of deception. Since deception based defense focuses on manipulating the adversary, it is essential to keep tracking of the attacker's behavior and observing the reaction of the attacker after deception is applied.

Deception by nature is a mental game between attackers and defenders. While defenders try to guide attackers into the wrong way,  an intelligent attacker could potentially suspect or even detect part of the deception scheme and make adjustment to their actions accordingly. By carefully monitoring feedbacks such as behavior changes of attackers, defenders can assess the cognitive state of the adversary and estimate the outcome of the current deception action.
\end{itemize}

By carefully evaluating feedbacks, the defenders will determine the outcome of deploying the current round of deception: (i) believed, when the deception scheme has successfully introduced biases to the attacker; (ii) suspected, when the attacker detects some abnormal signals and may not fully believe the intended fake information; and (iii) disbelieved, when the attacker identifies that the deception strategy is being used. This represents a total failure of the defender. The third scenario creates a challenging situation for the defender: the hard decision is whether to exit the deception scheme completely, or to start a brand new deception game. On the other hand, the defender's assessment on the deception outcome depends on the quantity and quality of feedbacks. When the feedbacks are lacking or have a high level of uncertainties, defenders can only make the best educated guess.

\subsection{Deception Schemes and Common Actions}\label{SubSec:Schemes}
Deception-based cyber defense relies on two main types of actions: information simulation and dissimulation. Information
dissimulation is commonly used to hide information. Common methods include masking, repackaging, and dazzling \cite{grazioli2003deceived}.

\begin{itemize}
\item Masking: Masking attempts to hide or erase crucial information from the target in order to escape detection.
Data masking techniques, such as shuffling, substitution and encryption, are commonly used to protect critical information such as personal identifiable data.

\item Repackaging: Repackaging refers to the transformation of key characteristics of the target so that they look irrelevant or different from the original, hoping that the attack's attention may be distracted away from the target. IP address hopping \cite{al2011toward} is a recent repackaging technique such that hosts on a network will constantly have different IP addresses and that network flows cannot be easily tracked by adversaries.

\item Dazzling: Dazzling blurs or obscures key elements of the target without removing them in order to confuse the target with other objects. For example, software obfuscation obscures critical sections of either source or running code and has been used as a common practice in defending against reverse engineering \cite{kuzurin2007concept,barak2001possibility}.
\end{itemize}

Information simulation involves mimicking, inventing, and decoying techniques \cite{bell1991cheating}. The objective of simulation is to create and use false information to distract and to mislead adversaries.

\begin{itemize}
\item Mimicking: Creation of a fake entity through imitation by duplicating key characteristics or identity of another real entity. It is one of the most widely used simulation methods. For example, a honeypot \cite{spitzner2003honeypots} can mimic a real web site.

\item Inventing: Creation of non-existent entities with key elements and key characteristics that look real. There are subtle differences between mimicking and inventing. While the primary requirement of mimicking is to create a fake entity that should look like the original entity, invention creates a new entity that looks realistic.

\item Decoying: Decoying has been a widely-used method among all simulation techniques. A decoy is normally an object that just looks or behaves like a genuine object. It is used to distract attention to something different from the real target. For example, a decoy web server can be used to attract attackers. Decoying, mimicking and inventing are often used together. A decoy server could be mimicking the function of a real server of an organization (i.e., a  honey pot), or it could be invented just for the purpose of distraction.

\end{itemize}

Both information simulation and dissimulation techniques are often combined in a real world deception scheme.

\section{Current Research and Applications based on Cyber Deception}\label{Sec:CurResearch}
\begin{figure}[h]
    \centering
    \includegraphics[width=\FigWidth]{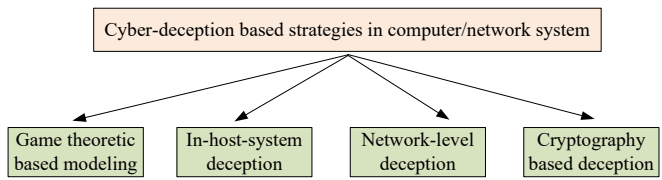}
    \caption{Categories of recent research results based on cyber deception.}
    \label{Fig:StrategyCat}
\end{figure}

Cyber attack and defense are an endless arms race. In a cyber deception game, both the defender and the attacker are trying to outsmart each other. In this section, we review recent techniques that leverage the idea of cyber deception to defend against potential intrusions. As shown in Fig.~\ref{Fig:StrategyCat}, current research can be summarized into four categories: game theoretic based modeling at the strategic level, network-level deception, host or device level deception, and cryptography-based deception. In the following, we review and summarize these four categories individually.

\subsection{Game-theoretic Modeling at the Strategic Level}
Game theory has been widely adopted at the strategy level to model the interactions between the defender and the attacker under varying security settings \cite{manshaei2013game, carroll2011game, 6426515}. For example, \cite{carroll2011game} modeled the defender-attacker interaction as a signaling game with non-cooperative two players. The research focused on establishing a dynamic game with incomplete information, and employed deceptive equilibrium strategies for network defense. In \cite{6426515},

a multistage Stackelberg game has been adopted for creating deceptive routing strategies in order to defeat jamming in multi-hop wireless networks. The deception framework models the scenario where the defender first deploys a proactive defense strategy, and then the adversary follows the protocol. Since the adversary is always resource limited, if the adversary wastes its limited resources, such as jamming power on the fake flow, the real data packets will have a much higher chance of arriving at the destination uninterrupted. Adopting a game model between the sending source node and the jammer helps the trade-off study between the deception cost and the impact, and allowed the defender to devise the most effective strategy against jamming.

More recently, the honeypot idea has been applied to Internet of Things (IoT) systems. A Bayesian game model is proposed in \cite{7442780} to represent the interactive deception process involving the attacker and the defender. Because both players
are willing to change their strategies upon learning from previous assessments, a repeated game model that enables the update of players' decisions has shown to be efficient under the Bayes setting.

\subsection{Network-Level Deception}
Under a typical network attack scenario, an adversary wants to gain control (or to disable) a valuable target in the network by first scanning the network, then hacking into a vulnerable system to gain access to more devices in the network from the initially compromised device. To prevent the adversary from successfully penetrating the network, the defender using cyber deception needs to engage with the attacker through the inspection of the incoming packets and try to identify the attacker's intent and capability, and then respond through a combined action of denying access to critical systems and staging misleading information to confuse attackers.

\begin{figure}[h]
    \centering
    \includegraphics[width=\FigWidthSmallEr]{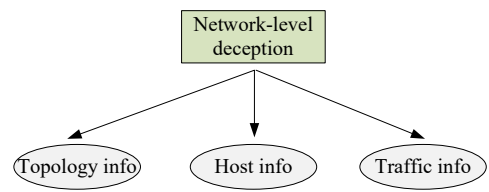}
    \caption{Network-level information can be potentially manipulated or fabricated to serve the deceptive purpose.}
    \label{Fig:NetLevelDeception}
\end{figure}

Several types of network-level information can be potentially manipulated or fabricated to serve the deceptive purpose, as shown in Fig.~\ref{Fig:NetLevelDeception}.

\begin{itemize}
\item Network topology information. The work in \cite{6735885} proposed using modified network topology response to deceive an adversary's traceroute probe, which has been used often by the adversary as the first step to discover the forward path of data packets.

Through the investigation of two strategies (random and intelligent deceptions), \cite{6735885} showed that the random deception scheme can add noise to distort the adversary's topology discovery; while the intelligent deception approach produces a believable, but incorrect network topology to mislead the adversary.

\item Network host information. In \cite{7346907}, a mobile honeypot system for industrial control systems was introduced. This system can be placed in many network locations to provide an additional layer of defense to disrupt attacker's reconnaissance activities. In addition the system can provide the defender with an early notification of incoming attacks.

Network tarpits \cite{tarpit} have been used as a form of defensive cyber deception to masquerade as many fake hosts as possible to deceive or to confuse network scanner. It is worth noting that in \cite{14abd-acsac}, an active probing detector Degreaser was developed to detect tarpits based on packet fingerprinting. Degreaser could be potentially used by an attacker as a detection tool. It is critical to revise current generation of tarpits to continue using the technique as a network security mechanism.

\item Network traffic information. The work in \cite{7346842} proposed an adaptive approach to deceive an attacker that actively collects traffic data in an attempt to obtain system fingerprints and to find a potential target. The proposed deceptive defense approach manipulates outgoing traffic so that it resembles traffic generated by a host with different system profiles (e.g., operating system and service).

The work in \cite{6750078} presented a new approach for self-configuring honeypots that can carry out passively check on network traffic of cyber-physical systems (CPS) and adjust to the sensing environment to create deceptive network entities that can be used to attract malicious players.

\end{itemize}

\subsection{In-host-system Deception}
In-host-system deception quite often lets the attacker enter a target system in a controlled manner. The deception setting in the target system can be used not only to mislead the attacker, but also help defenders to gather essential information about the attacker who has entered the system.

A typical example of in-host-system deception is honey patching proposed in \cite{araujo2014patches}, which reformulates traditional security patches into honey-patches that can confuse attackers by making it difficult for them to determine if a potential compromise has been made successfully or not. When a system detects an attempt to exploit, the honey patch redirects the attacker to an un-patched decoy where the attack is allowed to proceed. In the meantime, the decoy setting allows the defender to gather information about the attack and to potentially identify previously unseen malware. In addition, using decoy, misinformation can be presented to the attacker through falsified data or system settings.

Another recent work of in-host-system deception called honeywords \cite{juels2013honeywords} generates extra hashed fake passwords for each user's account in a system. Even when a file containing hashed passwords was compromised and the hash value could potentially be reversed, an adversary still cannot tell a real password from a similar-looking honeyword. This approach effectively adds another layer of defense to password file protection. In addition, when the adversary tries to use the cracked, but wrong password, the
system will instantly recognize the password hacking attempt.

\subsection{Cryptography based Approach}
Recently, a new cryptographic primitive known as honey encryption is introduced to help enhance system resilience against brute force attacks \cite{6876246, juels2014honey}. A honey-encrypted ciphertext has a unique property that an attacker could use a wrong key to yield a valid-looking output message, but the attacker cannot distinguish whether it is the correct plaintext or not. By adopting the honey encryption scheme, the defender can defeat brute-force attackers when they try to guess keys randomly. A recent application called Honey Chatting has been proposed in \cite{7472064} to combat eavesdropping attacks by applying honey encryption to chatting applications. Honey encryption ensures that an attacker will not be able to verify if a right key is used to decrypt a chatting message thus cannot determine the exact content of the message.

The application of honey encryption relies on a highly accurate distribution transforming encoder (DTE) over the message space. Unfortunately, the use of DTE severely impacts the practicality of honey encryption, mainly due to its inapplicability to more complicated structured data. Building an efficient and precise DTE is the main challenge when extending honey encryption into a varieties of practical applications. The work in \cite{huang2015genoguard} constructed an efficient DTE for genomic data that offers an information-theoretic security guarantee against message-recovery attacks.

\section{Major Research Issues Going Forward}\label{Sec:Directions}
Although the use of deception to enhance cyber defense has shown a number of interesting and promising results, it is still an under-explored area. There are many interesting research topics that await further investigation. In this section, we discuss key research issues that need to be addressed and identify steps going forward  towards the goal of establishing the scientific foundation of a full-fledged cyber deception framework for practical applications.

\subsection{Precise Adversarial Model}
Understanding adversaries is essential for a cyber deception scheme to succeed. However, obtaining a good understanding of adversaries has proven to be a huge challenge. A clear definition of the adversary model has long been desired. It will serve as a key basis for creating, analyzing and assessing a cyber defense technique \cite{cybenko2014adversarial}. We need a precise adversarial model for both creating a deception scheme and evaluating its effectiveness.

Current deception techniques often exploit cognitive biases of adversaries \cite{17wl-sp}. To be able to do that, defenders need to learn as much as possible about the cognitive state of an adversary which may include relevant information on knowledge, capability, intent, and decision process, and any biases from social/cultural factors and other surrounding issues. The cognitive state model is usually individualized, making it hard to generalize. However, a good human model for adversary cognitive state estimation is essential for deception design and deployment, because key elements of any successful deception scheme is to exploit biases of adversaries' cognitive state and leverage the weaknesses of their decision making process to our advantage.

Unlike MTD schemes which depend on creating computational complexity to increase attack difficulty and cost to defeat adversaries, deception relies on a better understanding of our opponents, especially their weaknesses and biases in knowledge and decision making process so that they could be exploited accordingly.

\subsection{Continuous, Multi-round Engagement}
As noted in Section~\ref{Sec:Models}, cyber deception can be considered as a two-party interactive game over time, where the defender must keep engaged with the attacker. However, it could be very challenging or sometimes even impossible to collect information about a potential attacker whom we have not been interacted with \cite{17wl-sp} before. As a result, people tend to think that cyber deception may not provide a good defense against zero-day attacks. However, this line of thinking may be faulty since to exploit zero-day attacks, adversaries still need to go through probing, learning and planning stages of their cyber kill chain. A key reason for having zero-day attacks is because the defender is unaware of the early actions taken by the attackers. Proactive strategies, such as cyber deception, aim to change that by emphasizing early engagement with adversaries. From that perspective, cyber deception schemes are extremely valuable for defending against future zero-day attacks because the first phase of cyber deception is to engage and gain the knowledge of an adversary as much as possible, and as early as possible.

To engage with adversaries during their reconnaissance phase, defenders can proactively craft honeypot- or honeynet-like systems that may attract adversaries' probing and leverage the interactions to learn more of their intent and their techniques.

Based on initial information learned, defenders can update their ``honey'' schemes to gain more knowledge. This suggests that the defender-attacker engagement can and should involve multiple rounds of interactions where defenders would adapt their system dynamically. The honey system can be adjusted based on initial knowledge on the adversaries. In the meantime, the system can start staging false information while being probed. This fits the thinking behind early work on game theoretic approaches to establish trade-off principles between truth disclosure and fake information projection \cite{carroll2011game, heckman2015denial} that may help guide honey system interactions and adversary engagement.

\subsection{Manipulation of Adversarial Mind}
The ultimate goal of cyber deception is to manipulate adversaries' decision process and to mislead them into wrong actions. Although manipulating or controlling a physical system has been studied extensively in the field of control theory \cite{bertsekas1995dynamic, bubnicki2005modern}, manipulation of human decision process is a much challenging process which has not been studied extensively. In prior psychology research, human mind manipulations through persuasion and influences have been studied \cite{hogan2010psychology}. However, these approaches are mostly through direct human interactions, including both verbal and non-verbal communications. In the cyber deception domain, the manipulation or influence is mostly through carefully planned cyber artifacts such as fake files, fake devices, and fake information.

In modern control theory \cite{bertsekas1995dynamic, bubnicki2005modern}, controllability and observability are two key concepts that have been used to determine whether a physical system can be measured and controlled, and how successful the control attempt will be. Observability refers to the ability to see and measure system attributes and to determine internal states, while controllability refers to the capability to change the target system to the desired state. The same two concepts may also apply to cyber deception and be used to quantify the ability to observe adversaries and the degree of manipulation that can be reached.

It is highly desirable to get as much observability of the deception target as possible, in order to apply deception schemes effectively to drive an adversary's mindset into the ``desired'' wrong state. The key research question still remains on how to formally define ``observability'' with regards to a human adversary's understanding and ``controllability'' that can be used to quantify the effectiveness of deception schemes. Initial research has touched upon integrating human factors in cyber security \cite{schultz2005human, bowen2011measuring} as well as using control theory to interpret human behavior \cite{carver2012attention}, but substantial new research needs to be carried out from the security and deception perspective with human factors in the loop, and with a possibility of leveraging insights gained from the rich set of modern control theory \cite{bertsekas1995dynamic, bubnicki2005modern} to interpret and measure cyber deception.

\subsection{Usability Analysis and Quantification}
Usability of an information system refers to the degree of ease, effectiveness, and efficiency that users can learn and use the intended functions provided by such a system \cite{nielsen2003usability}. It is one of the most important metrics that any information system designer needs to pay attention to, including proactive cyber systems. Initial study has shown that any security process needs to incorporate user experience into consideration in order for it to be successful \cite{cranor2014better}. When creating and deploying proactive techniques such as deception to thwart potential attacks, it is important to include usability in metrics to quantify the success of a deception scheme. Deception is meant to disrupt, distract and mislead adversaries through selected true and fake information setup and disclosure. However, a deception scheme needs to ensure that it will make the minimum impact to normal users. Key questions such as on how to prevent average users from being confused or even disrupted by deception information are largely untouched.

Attack surface quantification \cite{manadhata2011attack} provides us with a good measurement of MTD effectiveness. However, metrics for deception are yet to be defined. It is our belief that any deception effectiveness measurement should also incorporate usability impact.

\subsection{Combining Deception and MTD Approaches}

Cyber deception and MTD schemes have received lots of research interests recently. There were over 40 MTD schemes  based on a 2013 survey report \cite{okhravi2013survey}, and the number of scheme is much higher today. In the meantime, deception schemes are gaining popularity and often combined with MTD as a complementary method. When MTD is adopted to increase diversity and complexity of a target system, deception can augment MTD by adding fake setups for better defense. While MTD makes it harder and more expensive for adversaries to observe and attack, deception will distract and mislead in the meantime.

To disrupt the adversaries' cyber kill chain, both schemes can be used at the same time or at different stages. For example, recent works on network anti-inference \cite{15lw-info,15lww-milcom} suggest combining both MTD technique (i.e., dynamic change of network routing) and deception technique (i.e., adding fake network traffic) to effectively disrupt adversaries' capability to identify network flows and to determine critical nodes of the network. In addition, both techniques can be applied at different stages: while both MTD deception can mislead adversaries' reconnaissance efforts, MTD can further increase the complexity and difficulty for them to penetrate a system. More research is needed on how to combine both methods seamlessly and to maximize protection.

\section{Conclusion}\label{Sec:Con}

In this paper, we reviewed emerging cyber deception research from high-level concepts to detailed techniques developed recently for protecting computer and network systems. Cyber deception has shown its promising use but is still in its infancy with many challenging issues to be addressed. More research efforts will be needed to lay a solid scientific foundation that spans multiple disciplines including computer and network security, cryptography, control theory, and cognitive science.

\bibliographystyle{IEEEtran}
\bibliography{deceptionsurvey}

\end{document}